\documentclass[10pt,prb,sort,twocolumn]{revtex4}
\usepackage{amsmath,amssymb}
\usepackage{citesort,epsfig}
\begin{document}

\title{Solution stability and variability in a simple
model of globular proteins}
\author{ Richard P. Sear}
\affiliation{Department of Physics, University of Surrey,
Guildford, Surrey GU2 7XH, United Kingdom
{\tt r.sear@surrey.ac.uk}}

\begin{abstract}
It is well known amongst molecular biologists that proteins
with a common ancestor and that perform the same function in
similar organisms, can have rather different amino-acid sequences.
Mutations have altered the amino-acid sequences without
affecting the function.
A simple model of a protein in which the interactions are
encoded by sequences of bits is introduced, and used to study how
mutations can change these bits, and hence the interactions,
while maintaining the
stability of the protein solution. This stability is a simple
minimal requirement on our model proteins which mimics part
of the requirement on a real protein to be functional.
The properties of our model protein, such as its
second virial coefficient, are found to vary significantly from one model
protein to another. It is suggested that this may also 
be the case for real proteins in vivo.
\end{abstract}

\maketitle

\section{Introduction}

Proteins are linear heteropolymers: they are linear sequences of
monomers, each of which is one of twenty different types.
Different proteins have different sequences of amino acids. 
These differences allow proteins to perform the huge range of tasks
they do in living cells.
But this does not mean that 2 proteins
that do the same job necessarily have the same sequence.
For example, many organisms have enzymes called adenylate
kinases which perform essentially the same job in the cytoplasm
of each organism. But the amino acid sequences of adenylate
kinases vary very widely, even though they are all doing the same
job in more-or-less the same milieu. Below are the amino-acid
sequences of the adenylate kinases of two prokaryotes
\cite{proteins}.
First that of {\it Escherichia~coli}
\[
{\small
\begin{array}{l}
\mbox{\tt MRIILLGAPGAGKGTQAQFIMEKYGIPQISTGDMLRAAVKSGSELGKQAK}\\
\mbox{\tt DIMDAGKLVTDELVIALVKERIAQEDCRNGFLLDGFPRTIPQADAMKEAG}\\
\mbox{\tt INVDYVLEFDVPDELIVDRIVGRRVHAPSGRVYHVKFNPPKVEGKDDVTG}\\
\mbox{\tt EELTTRKDDQEETVRKRLVEYHQMTAPLIGYYSKEAEAGNTKYAKVDGTK}\\
\mbox{\tt PVAEVRADLEKILG}\\
\end{array}
}
\]
and secondly that of {\it Vibrio~cholerae}
\[
\begin{small}
\begin{array}{l}
\mbox{\tt MRIILLGAPGAGKGTQAQFIMEKFGIPQISTGDMLRAAIKAGTELGKQAK}\\
\mbox{\tt AVIDAGQLVSDDIILGLIKERIAQADCEKGFLLDGFPRTIPQADGLKEMG}\\
\mbox{\tt INVDYVIEFDVADDVIVERMAGRRAHLPSGRTYHVVYNPPKVEGKDDVTG}\\
\mbox{\tt EDLVIREDDKEETVRARLNVYHTQTAPLIEYYGKEAAAGKTQYLKFDGTK}\\
\mbox{\tt QVSEVSADIAKALA}
\end{array}
\end{small}
\]
\noindent
where the sequences are given as a sequence of the 1-letter codes
for the amino acids of which they are made. The first amino acid
is an M (Methionine), the second
is an R (Arginine) and so on.
The sequence is read as English text, from top left to bottom right.
See any molecular biology or biochemistry textbook
\cite{thecell,stryer,voet} for an introduction to amino acids
and proteins.
Note that there are many differences between
the sequences!  The
amino-acid sequences of proteins are very different while
keeping the function. Also, we picked adenylate kinases only in order to
have a concrete example, it is a general property of proteins.
The
function of adenylate kinases is irrelevant to our discussion of stability,
beyond the fact that they function as enzymes as monomers
in solution inside cells. Here we will concentrate entirely
on globular proteins, the proteins that exist in solution not
embedded in membranes.

\begin{figure}
\begin{center}
\caption{
\lineskip 2pt
\lineskiplimit 2pt
Schematic representation of a model protein,
with the 3 visible patches represented by `barcodes': a sequence
of stripes, light for hydrophilic and dark for hydrophobic.
The model shown has $n_B=4$ bits of which 2 are hydrophobic (0)
and 2 are hydrophilic (1) in each case.
For example, the `barcode' of the front patch is $0101$.
\label{model}
}
\vspace*{0.3in}
\epsfig{file=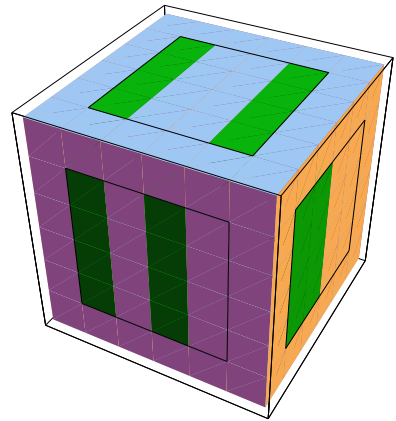,width=2.0in}
\end{center}
\end{figure}

Now, the simplest thing to do when faced with this radical
difference in sequence without a corresponding difference in function
is to ignore it. To assume that the 2 proteins interact and behave
in a very similar manner. But do they? As they both function as
proteins inside the cytoplasm of bacteria they are both clearly
soluble and do not stick to things they should not stick to
{\em in vivo}. However, this does not mean that their solubilities,
for example, are necessarily equal. Both their solubilities are
sufficient to allow them to function but one may exceed the
minimum solubility by a large margin and one by a small margin.
It would be of interest to know what these margins are and how they
vary from protein to protein, not only because we wish to understand
how proteins function
and have evolved {\em in vivo}, but to help us process, purify,
and crystallise proteins. If a protein is only marginally soluble
in the conditions {\em in vivo} then it may aggregate when
its environment (salt concentrations, temperature etc.) are altered.
We would like to understand  and to be able to predict,
the variability of properties, such
as solubility, of proteins.

We will focus on the stability of solutions of proteins in their native state,
i.e., we assume that the protein has folded into its native state and
remains there. Thus we consider only folded proteins sticking together
due to their surfaces attracting each other, not proteins partially
unfolding and then aggregating due to the hydrophobic regions of the
protein exposed by unfolding, attracting each other. So, our proteins
will always be compact objects, more like colloidal particles than
conventional polymers. This allows us to avoid the complex problem
of protein (un)folding. Effectively, we assume that proteins such
as the adenylate kinases of {\it E.~coli} and
{\it V.~cholerae} differ only in their surfaces. Replacing
one surface amino acid in the chain by another then changes only the
surface and through that the protein-protein interaction. If a
hydrophobic amino acid replaces a hydrophilic amino acid in a position
on the chain where the chain is at the protein's surface, then we expect
the surface to become more sticky, which would tend to decrease
the second virial coefficient, whereas replacing a hydrophobic amino
acid by a hydrophilic one should have the opposite effect.
For simplicity, instead of having 20 different types of amino acids
at the surface, we use a model whose surface is described by bits
which have only 2 values: hydrophobic and hydrophilic. This is a rather
gross approximation, the amino acids vary widely in size, some are charged,
but we want the simplest possible model. The model is an extension
of that considered in Ref.~\onlinecite{sear02}. A protein molecule
is modeled by a cube, whose 6 faces interact with a short-ranged
attraction, which is here determined by a sequence of $n_B$ bits.
In Ref.~\onlinecite{sear02} the interaction between faces was taken
to be a random variable; we will discuss the differences between
that model and the more complex one considered here in the conclusion.
A schematic of the model is shown in Fig.~\ref{model}.

We have talked of our model proteins being soluble {\em in vivo}.
Real proteins have evolved to be so. The cytoplasm of bacteria
such as {\it E.~coli} and
{\it V.~cholerae} is very complex: bacteria typically have
a few thousand different proteins \cite{vibnote},
and any one of these proteins
is then surrounded by thousands of different proteins, as well as
RNA and DNA, small molecules such as nucleotides etc..
An individual enzyme must be soluble in the sense that it does not stick
too strongly to not only other proteins of the same type but those of
all the other types, as well as not binding to the RNA, DNA, etc..
In future work, we will address this problem, but here we will keep
things simple and consider only interactions between model proteins
of one type. We will calculate the second virial coefficient only
for the interaction of 2 model protein molecules of the same type.
This is not realistic for a enzyme in a bacterial cell as an individual enzyme
will be present at rather low concentrations, even though the
total protein concentration in bacterial cells is around
20\% by volume \cite{neidhardt,thecell}.
It is however, a good place to start, and is realistic for a few
exceptional cells, such as our red blood cells which contain very
high concentrations of a single protein: hemoglobin. Future work
will address this issue and
will also look at proteins which bind to other proteins, as many
proteins do.

\begin{figure}
\begin{center}
\caption{
\lineskip 2pt
\lineskiplimit 2pt
A schematic of protein space for both a protein and a model
protein. The arrows represent mutations changing a protein
located at one point in protein space into a neighbouring protein.
\label{protsp}
}
\vspace*{0.1in}
\epsfig{file=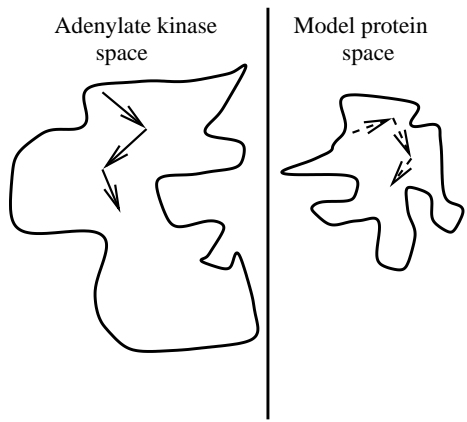,width=2.4in}
\end{center}
\end{figure}

The proteins whose sequences we gave
in the first paragraph are presumably
orthologs: they are both descended from a common ancestral
protein but have evolved independently, keeping their
function the same, since the {\em E.~coli} and {\em V.~cholerae}
lineages separated.
The fact that proteins with the same function, but that have
evolved independently in different species, can have very
different amino-acid sequences, is well known.
Paralogs, proteins created by duplication of a gene, also
start with identical sequences but have sequences that diverge with time.
The differences
are believed to have
arisen via random mutations which are not rejected by natural
selection because they are not actively deleterious (to the survival
of the organism) but also do not have any selective advantage.
This theory of mutations changing the amino-acid sequences
of proteins without improving or reducing its ability to function
is called the theory of neutral evolution
\cite{kimura68,king69,ayala97}.
The constraints placed on this neutral evolution by
the requirement on the protein to fold have been
considered \cite{bastolla00,bastolla02,deeds03,tang00,bornberg99,taverna02},
but not those due to the requirement of the protein to be soluble.
The constraints placed on the sequences of RNA by the requirement
to be functional and the evolution of these sequences, are analogous
to the constraints on the sequences of, and evolution of proteins.
They have been extensively studied and in many respects are rather
better understood, essentially because RNA is simpler than protein.
See the review of Higgs\cite{higgs}.
However, there has been some work which has considered protein-to-protein
variability \cite{sear02,sear03},
see also Ref.~\onlinecite{rosenbaum99}

We will generate our model proteins at random (subject to the solubility
constraint) and assume that neutral evolution of proteins
is close to a random walk from one sequence to another.
This random walk occurs in what is often called `protein space'
\cite{smith70}, with each sequence a unique
point in this space and 2 sequences neighbours if 1 of them
can be transformed into the other by a single mutation.
This protein space is vast.
The set of soluble proteins exists in this protein space
as a set of points, 1 for each soluble protein.
A schematic of the protein
spaces of proteins and model proteins is shown in Fig.~\ref{protsp}.
It is only very schematic, the space is huge and many dimensional.
In each case the arrows represent
a single mutation changing a protein into a neighbouring protein.
Below, we will generate random walks for
our model proteins, and these will sample all soluble states
with equal probability. When we come to applying our results
to real, not model, proteins, we will have to assume that neutral
evolution also samples proteins which are soluble with reasonably
uniform probabilities.

In the following section, we will perform a simple analysis of
sequence data, to look at variations in the number of hydrophobic
amino acids. 
The model is defined in section \ref{secmodel}, and the stability of its
solutions estimated and discussed in section \ref{secsol}.
The last section is
a conclusion.

\section{Analysis of sequence data}

The sequences of the adenylate kinases of  
{\it E.~coli} and {\it V.~cholerae} are both of
viable enzymes, they are soluble
{\it in vivo} and catalyse a reaction. Looking at them,
an obvious question to ask
is: How many sequences of amino acids are there, that fold
up to form viable adenylate kinases?
Both adenylate kinases have 214 amino acids.
As there are 20 types of amino acids there are
$20^{214}\simeq 10^{278}$ different amino acid sequences of 214
amino acids. An enormous number, of which presumably the vast majority
do not fold into a unique native state,
let alone are soluble and act
as a catalyst. But it seems likely that the number of possible
amino acid sequences that correspond to viable adenylate kinases
is huge.

\begin{figure}
\caption{
\lineskip 2pt
\lineskiplimit 2pt
A scatter plot of the fraction of its
amino acids which are hydrophobic, $h$, versus the number
of amino acids $M$. Results for the prokaryote members of the
family of adenylate kinases are shown. The PROSITE
accession number is PS00113.
\label{fhbac}
}
\vspace*{0.3in}
\begin{center}
\epsfig{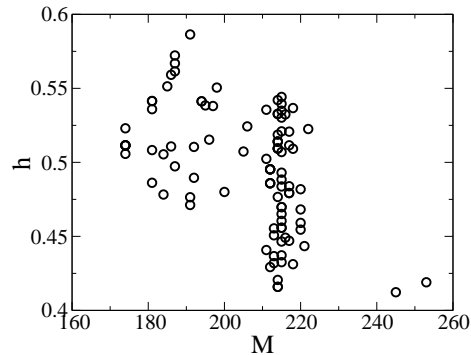}
\end{center}
\vspace*{0.2in}
\end{figure}

A database at SWISSPROT \cite{swissprot,spref},
called PROSITE \cite{prosite,prositeref},
identified 152 amino-acid sequences as belonging to the adenylate-kinase
family of proteins (PROSITE accession number PS00113).
It did so by locating the amino acids of the
active site of an adenylate kinase \cite{prosite,prositeref}.
104 of these sequences
are from prokaryotes, of which we eliminate 4 sequences as they
contain less than 100 amino acids and are presumably not complete
proteins. This leaves 100 adenylate kinases; 2 of these kinases
are the ones whose sequences are in the first paragraph.
We can calculate
the fraction of the amino acids of these adenylate
kinases that are hydrophobic, $h$,
and plot this against the length of the sequence, $M$:
the total number of amino acids in the sequence. The
results are shown as a scatter plot, Fig.~\ref{fhbac}.
The 9 amino
acids G, A, V, L, I, M, P, F and W, are taken to be hydrophobic,
and the remaining 11 to be hydrophilic. Here
each amino acid is represented by its 1-letter code: G for glycine,
A for alanine, etc.. The 9 hydrophobic amino acids are those
whose side chains are classified as nonpolar in Ref.~\onlinecite{voet}
(Table 4-1, p58). There is some arbitrariness in where the
dividing line is drawn between hydrophobic and hydrophilic amino
acids, but different dividing lines give rather similar spreads
in $h$.

For the present work, the key observation is that the fraction
of an adenylate kinase's amino acids which are hydrophobic
varies from protein to protein, as do other properties
such as their net charge \cite{sear03,unpub}.
In section \ref{secsol} we will find that for our 
model, with a constraint imposed that
model proteins are soluble, there is
scatter in the fraction of its bits that are hydrophobic.

\section{Model}
\label{secmodel}

The model is chosen to be as simple and as generic as possible,
while having interactions which are
mediated by surface patches whose interactions
are a function of sequences or string of bits.
The protein-protein interactions then depend on
the values of these bits, some sets of values give proteins which strongly
attract each other while other sets give proteins which largely
repel each other. This is perhaps the simplest model of a globular protein
which allows for mutations.
Within the model these mutations flip one of the bits, a model
of a mutation which converts a surface residue from
a hydrophobic amino acid to a hydrophilic amino acid, or vice versa.
A schematic of the model is shown in Fig.~\ref{model}. An amino acid
of a protein is called a residue.

The model protein is a cube, with each of its 6 faces
having a single patch \cite{sear02}.
The lattice
is cubic and each protein occupies 8 lattice sites arranged
2 by 2 by 2,
see Fig.~\ref{model}. We make the model 2 sites across to reduce the
range of the attraction, which is 1 site, to half the diameter
of the hard core.
The model proteins can rotate, and so
have 24 distinct orientations. Each of the 6 faces of the cube
has a patch, labeled $i=1$ to 6, with patches 1 to 4 clockwise around
a loop of 4 of the faces, and patches 5 and 6 on the remaining
2 faces.
The interactions between
model proteins are pairwise additive and consist of 2 parts.
The first is simply an excluded-volume interaction:
2 proteins cannot overlap.
The second is that if the faces of 2 proteins
are in contact
there is an energy of interaction between the 2 touching patches of the
2 proteins. By in contact we mean that
the faces must
overlap completely otherwise the energy of interaction is taken to be zero.
Also, the model is such that the energy of interaction between two
touching patches is a constant which does not change when the two
proteins are rotated about the axis joining their centres.
The touching patches are those on the faces of the 2 proteins
that face each other. This is all as in Ref.~\onlinecite{sear02}, the
difference is in how the interaction energy of a pair
of patches $i$ and $j$, $u_{ij}$, is specified.

How a patch interacts
is specified by a
sequence or string of $n_B$ bits. If a bit has a value of 1 then
the bit is said to be hydrophilic or polar, whereas if it has a value of 0
then it is hydrophobic. The interaction energy of a pair
of touching patches, $i$ and $j$, is then given by
\begin{equation}
u_{ij}=-\epsilon\sum_{\alpha=1}^{n_B}\left(b^{(i)}_{\alpha}-1\right)
\left(b^{(j)}_{1+n_B-\alpha}-1\right),
\end{equation}
where $b^{(i)}_{\alpha}$ is bit number $\alpha$ of patch $i$.
$\epsilon$ is the interaction energy of 2 hydrophobic bits.
We use energy units such that the thermal energy $k_BT=1$.
Thus to calculate the interaction the string of bits
of 1 of the patches is reversed and then the energy is just the
sum of the number of pairs of corresponding bits where both
bits are 0, are hydrophobic. The only interaction is between 2 hydrophobic
bits; there is no hydrophobic-hydrophilic or
hydrophilic-hydrophilic interaction.
The reason one of the strings is reversed is that if this is not
done then the interaction between like patches, $j=i$, is just
$\epsilon$ times the number of 0s in $i$'s string. Reversing
the strings removes this problem in a simple way.
Of course, the interactions form a symmetric square matrix,
$u_{ij}=u_{ji}$. Each of the 6 patches is taken to be labeled
and so distinguishable, i.e., we take a pair of proteins where one
protein can be obtained from the other by swapping a pair of
the strings of bits, as 2 different proteins.

Thus, a protein is specified by giving values to the $6$ strings
of $n_B$ bits, and so there are $2^{6n_B}$ possible different
proteins. For all but rather small values of $n_B$, this is
a very large number of possible proteins, e.g., for $n_B=18$, we
have $3\times 10^{32}$ different model proteins. This is however, much
smaller than the number of possible real proteins.
Most of the calculations have been done for $n_B=18$, with a few
for $n_B=12$, for comparison. We choose $n_B=18$ as being
a sensible number as then the total number of bits which
describe the surface is 108. Adenylate kinases, for example,
have around 200 amino acids, of which about half are on the surface.
Thus, we have about 1 bit per surface amino acid.
Our model proteins can be thought of
as existing in `protein space' with each possible protein represented
by a point in this space, and each protein has $6n_B$ neighbours,
each of which is obtained by flipping 1 of the bits of the protein,
see Fig.~\ref{protsp}.

The second virial coefficient $B_2$ of our lattice model is given by
\cite{sear02}
\begin{equation}
B_2=\frac{1}{2}\left[27-\frac{1}{6}
\sum_{i=1}^6\sum_{j=1}^6\left(\exp\left(u_{ij}\right)-1\right)\right],
\label{b2def}
\end{equation}
where the first term inside the brackets comes from excluded-volume
interactions and the second from the interactions between touching patches.
The number 27 comes from the fact that each model protein excludes
other proteins from a cube of 3 by 3 by 3 lattice sites. Thus,
in the high temperature limit
$B_2=B_{2hc}=27/2$.
The sums over 24 orientations reduce to sums over 6 orientations
as rotating either of the 2 molecules around the axis joining
their centres does not change the energy. The factor in front
of the double sum is a normalisation factor of $1/36$ times the
6 possible lattice sites that one molecule can occupy and be
adjacent to the other molecule.

\section{Stability of solutions}
\label{secsol}

Unless $\epsilon$ is small, many of the $2^{6n_B}$ model proteins
strongly attract each other leading to condensation,
gelation, and possibly crystallisation.
By condensation we mean
the formation of coexisting dilute and concentrated protein
solutions, as have been studied extensively for the protein
lysozyme \cite{muschol97,vliegenthart00,rosenbaum99}.
Only a fraction of the model proteins are viable in the sense that they
are stable as single phase solutions. Clearly proteins cannot
condense {\em in vivo} without severely impairing the organism's function.

The attractions affect the phase behaviour through and can be measured
by, the second-virial coefficient.
In the absence of attractions the second-virial coefficient
is approximately 4 times
the volume of a particle (assuming the particle is not too anisotropic).
Attractions decrease its value until eventually the pressure
does not increase monotonically but decreases over a range of densities
due to the negative virial coefficient; a van der Waals loop forms.
If we impose the constraint
that the second virial coefficient
be above a certain value, where we believe
the pressure will be a monotonic function of density,
we can quantify what fraction
of our model proteins satisfy this constraint and so have solutions
which are stable.

\begin{figure}
\begin{center}
\caption{
\lineskip 2pt
\lineskiplimit 2pt
A plot of the fraction of proteins with stable solutions, $f_v$,
as a function of $\epsilon$. The solid
and dashed curves are for $n_B=12$, and 18 bits, respectively.
\label{fvia}
}
\vspace*{0.3in}
\epsfig{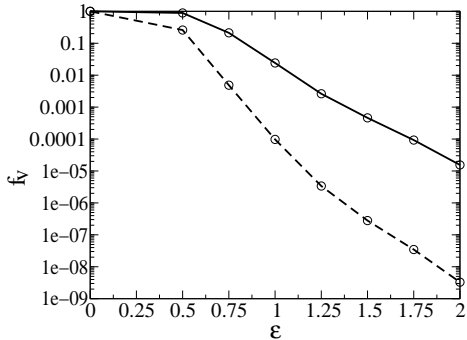}
\end{center}
\end{figure}

We insist that the reduced second virial coefficient
satisfy $B_2/B_{2hc}\ge -1$, in order for the protein 
to be viable. The fraction of proteins which
are viable, according to this criterion, is denoted by $f_v$. It
is determined by generating proteins at random, setting each bit
to be 0 or 1 with equal probabilities, and finding the
fraction with $B_2/B_{2hc}\ge -1$.
See the appendix for further details of the computations.
The value of $B_2/B_{2hc}$ at the critical point, the highest point
on the curve separating the 1 and 2-phase regions of a phase
transition into coexisting solutions, is typically a little less than $-1$,
unless the attraction is very anisotropic. For the canonical model,
hard spheres plus a long-range attraction,
the critical point occurs when $B_2/B_{2hc}=-1.65$, and
provided the attraction remains isotropic this value changes
little even if the attraction is made quite short ranged
\cite{vliegenthart00}. If the attraction is very anisotropic
then $B_2/B_{2hc}$ can (depending a little on the precise nature
of the anisotropy) be much more negative at the critical point
\cite{sear99,curtis01,kern03}, but for simplicity we insist on 
$B_2/B_{2hc}$ being above a fixed value for all our proteins,
regardless of how anisotropic their attractions are.
Crystallisation out of not-too-concentrated solutions also
requires as a minimum,
attractions of about the strength required to make $B_2/B_{2hc}$ around $-1$.
The propensity to crystallise depends on the details of the
attraction, for work on the earlier version of this model
with random values of the patch-patch attractions, see Ref.~\onlinecite{sear02}.

Results are shown, as a function of $\epsilon$, for $n_B=12$ and 18,
in Fig.~\ref{fvia}. As might have been expected, as
$\epsilon$ increases, the fraction of viable proteins
decreases exponentially, but note that even for $n_B=12$ and
$\epsilon=2$, there are still $7.1\times 10^9$
viable proteins, a very large number.
Partly, what is happening is that as $\epsilon$ increases then
fewer and fewer hydrophobic bits are allowed, and as the fraction of bits
that are hydrophobic decreases, then the number of possible proteins
decreases: there are many possible proteins with close to half
their bits 0s and half 1s, but only one with all its bits equal to 1.
Partly, what happens is that correlations are introduced between
the hydrophobic bits in the strings. The hydrophobic bits
tend to avoid each other,
e.g., if all 6 strings have all their bits from 1 to $n_B/2$
(assuming $n_B$ is even) hydrophilic, then any or all of their
bits from $n_B/2+1$ to $n_B$ may be hydrophobic without there
being any attractions. Thus here the hydrophobic
bits avoid each other, in order to avoid the
attractive interactions which make the second virial
coefficient negative and thus violate our solubility condition.

\begin{figure}
\begin{center}
\caption{
\lineskip 2pt
\lineskiplimit 2pt
\label{phobc}
A plot of the mean fraction of bits which are hydrophobic,
$\langle h\rangle$, the solid curve, and
of a measure of the correlation between a bit and the other bit
it interacts with, $\langle hh_p\rangle$, the dashed curve.
}
\vspace*{0.3in}
\epsfig{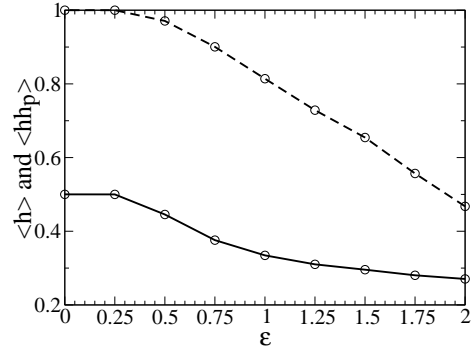}
\end{center}
\end{figure}

We can measure both these effects by defining 2 quantities. The first
is the mean fraction of bits which are 0, are hydrophobic. Denoting
this by $\langle h \rangle$, it is defined by
\begin{equation}
\langle h\rangle=\frac{1}{6n_B}\langle\sum_{i=1}^6\sum_{\alpha=1}^{n_B}
\left(1-b^{(i)}_{\alpha}\right)\rangle.
\label{hdef}
\end{equation}
The average denoted by $\langle\rangle$ is
over proteins which satisfy
our criterion for the stability of the solution.
We use $h$ to denote both the fraction of bits in our model
proteins that are hydrophobic, and the fraction of residues in
real proteins that are hydrophobic. A measure of
the correlation between the probability that a bit $\alpha$ is
hydrophobic, and that the bit $1+n_B-\alpha$
with which it interacts is also
hydrophobic is denoted by $\langle hh_p\rangle $, and is defined by
\begin{equation}
\langle hh_p\rangle=\frac{1}{36n_B\langle h\rangle^{2}}\langle
\sum_{i=1}^6\sum_{j=1}^6\sum_{\alpha=1}^{n_B}
\left(1-b^{(i)}_{\alpha}\right)
\left(1-b^{(j)}_{1+n_B-\alpha}\right)\rangle .
\end{equation}

We have plotted both quantities in Fig.~\ref{phobc}. The model
has $n_B=18$ bits and the quantities are plotted as a function
of $\epsilon$. As $\epsilon$ increases, the fraction of bits
which can be hydrophobic without the second virial coefficient
becoming too negative decreases. Also, the anticorrelations between
a bit being hydrophobic and the bit with which it interacts being
also hydrophobic increases. If there were no correlation between the
states of the 2 bits then $\langle hh_p\rangle=1$, which is true for
$\epsilon=0$, but this function decreases as $\epsilon$ increases.
If a bit is hydrophobic the bit with which it interacts is less likely
to be hydrophobic.
We have shown results just for $n_B=18$ but results
for other numbers of bits are similar.


\begin{figure}
\begin{center}
\caption{
\lineskip 2pt
\lineskiplimit 2pt
\label{b2}
The probability distribution function, $P$, for the reduced
second virial coefficient, $B_2/B_{2hc}$, for $n_B=18$ and
$\epsilon=2$.
}
\vspace*{0.3in}
\epsfig{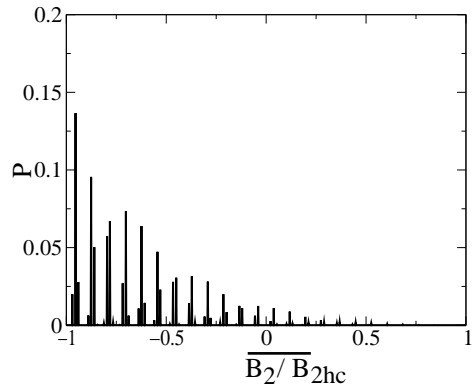}
\end{center}
\end{figure}

We only constrain the second virial coefficient to be above
a certain value, we do not constrain its precise value.
As the second virial coefficient is a function of the number
of hydrophobic bits on its 6 faces and as this number is an integer
between 0 and $n_B$, the second virial coefficient can only
take one of a set of values, and so the probability density function
for $B_2/B_{2hc}$ is a set of delta functions. We have plotted these
as spikes, with the height of each spike set to the probability
that $B_2/B_{2hc}$ has this value. We can see that the most likely
values of the reduced second virial coefficient are
near the minimum allowed value of $-1$.
This is simply because there are many more sets of strings with close
to half the bits hydrophobic than there are with most of the bits hydrophilic,
and the proteins with close to half the bits hydrophobic have very large
and negative second virial coefficients. There is only one protein
with all 108 bits hydrophilic but the number of proteins which
have 9 hydrophobic and 9 hydrophilic bits on each face is
$(18!/9!^2)^6\sim 10^{28}$. The probability distribution function for
{\em all} possible proteins (including those with $B_{2}/B_{2hc}<-1$)
is sharply peaked at a value much less than 1, for $n_B=18$ and
$\epsilon=2$, and Fig.~\ref{b2} shows just the high $B_2$
tail of this distribution.

\begin{figure}
\begin{center}
\caption{
\lineskip 2pt
\lineskiplimit 2pt
The probability distribution function, $P$, for the fraction
of bits hydrophobic, $h$. For $n_B=18$ and
$\epsilon=2$.
\label{fh}
}
\vspace*{0.3in}
\epsfig{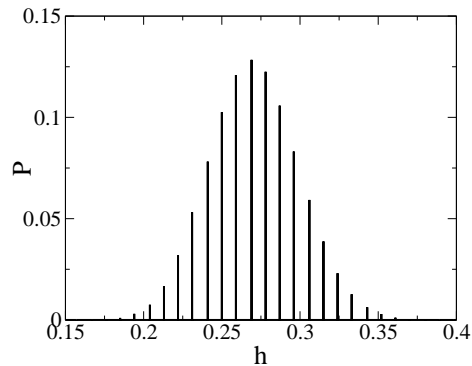}
\end{center}
\end{figure}

The probability distribution function, again
a sum of delta functions, of $h$ the fraction of hydrophobic bits,
is plotted in
Fig.~\ref{fh}. As with Fig.~\ref{b2}, $n_B=18$ and $\epsilon=2$.
The distribution is peaked at $h$ a little above $0.25$:
the mean value $\langle h\rangle=0.27$ and the standard deviation is $0.029$.
As $h$ increases towards $0.5$ then there are many more possible
proteins but a rapidly increasing fraction of these are not soluble
as a single phase
according to our criterion. Thus there is a trade off between the
number of possible proteins and the fraction that are soluble.
This trade-off results in most proteins having between 20\%
and 35\% of their bits hydrophobic. This of course depends on
$n_B$ and $\epsilon$. Increasing either one decreases $\langle h\rangle$
but the picture remains qualitatively the same.

Let us return to our results for adenylate kinases, Fig.~\ref{fhbac}.
Although it should be borne in mind that
many of adenylate kinases' hydrophobic amino acids will be in the
centre of the protein, not at its surface, we can still quantify
the scatter in $h$ for the kinases, and compare it to the scatter
in $h$ for the model proteins. But of course any comparison will be purely
qualitative.
The adenylate kinases have around 200
amino acids in total, of which about 100
are classified as hydrophobic.
We can try to model the
distribution functions for $h$,
for both adenylate kinases and our model proteins, with
\begin{equation}
h=n^{-1}\sum_{i=1}^n \zeta_i,
\label{random}
\end{equation}
where for an adenylate kinase
the sum is over all its amino acids in a protein, $n=M$,
and for a model protein the sum is over the
$n=6n_B$ bits. The $\zeta_i$ are independent random
variables which are 1 with probability $\langle h\rangle$ and zero otherwise.
For adenylate kinases, see Fig.~\ref{fhbac},
we find that the standard deviation of $h$ is $0.040$,
and Eq.~\eqref{random} gives a standard deviation of $0.035$, only a
little lower.
To obtain the value of $0.040$ we took the sum over 206 terms; 206 is the
mean length of the adenylate kinases in Fig.~\ref{fhbac}. Taking
all the proteins to be the same length will decrease the spread
slightly.
Note that we can predict the distribution of the
proteins' hydrophobicity reasonably accurately
using only the central limit theorem.

For our model proteins the standard deviation of $h$ is
$0.029$, while Eq.~\eqref{random} predicts $0.043$, which is rather
larger but still comparable.
Also, of course the shape
of the distribution in Fig.~\ref{fh} is quite close to Gaussian.
Thus, the results for our model proteins
are similar to those for real proteins, but as both are within
a factor of $1.5$ of a simple prediction based on assuming the
hydrophobic amino acids/bits are randomly distributed, it is
hard to draw definite conclusions from this.
The distribution of net charges can also be modeled assuming
that the charged amino acids are distributed at random
\cite{sear03,unpub}.


\section{Conclusion}

We started with the idea that globular proteins needed to be soluble
to function, and that their interactions depended on their surfaces
which in turn were sensitive to which types of amino acids
were at the surfaces of proteins. Then we defined a very simple
model of a protein, whose surface-mediated-interactions depended
on the values of strings of bits. A mutation in a protein
such as an adenylate kinase which
substituted a hydrophobic amino acid at the surface for a hydrophilic
one could then be modeled by flipping one of these bits.
Within our model, and with the constraint that a solution 
of the model protein
is stable; the second
virial coefficient is rather variable, its probability distribution
function is plotted in Fig.~\ref{b2}. The criterion
for the solution to be stable as a single phase
is taken to be that the reduced second virial coefficient
$B_2/B_{2hc}\ge -1$, which is enough for almost all fluids
to be above their critical point. The condition that the protein
solution be stable as a single phase is clearly a necessary
condition, although in fact the second virial coefficient may be
be more tightly constrained than this.
Although the model used is
simple, this variability does give credence to the idea
that the variation in the fraction of hydrophobic
amino acids in enzymes like adenylate kinase, see Fig.~\ref{fh},
gives rise to variability in the protein-protein interactions
of these enzymes. In other words, that the second virial coefficients
of {\em E.~coli}'s and {\em V.~cholerae}'s
adenylate kinases may be significantly different, even
though there is no obvious functional reason
why their physical properties should differ.
Unfortunately, virial coefficient measurements have not been
performed for families of proteins.
The variability is relevant to problems such as the purification
and crystallisation of proteins.
The separation of one protein from all the others
in an extract from
a cell which might contain thousands of proteins relies
on {\em differences} in physical properties, charge, surface stickiness, etc.,
between proteins.


The probability distribution function of the second-virial coefficient,
Fig.~\ref{b2}, is just the high $B_2$ tail of the distribution
of all proteins. The remainder of the distribution function is cutoff
by the requirement that $B_2/B_{2hc}\ge-1$. This full distribution function
has a peak at an $\epsilon$ dependent value of $B_2$; here
well below $-1$. Thus,
without the cutoff at $B_2/B_{2hc}=-1$, the distribution function is
similar to the Gaussian distribution function found for the earlier
model in which the patch-patch interactions were taken to be random
variables \cite{sear02}.
If we had kept with the previous model of describing with
random variables the patch-patch interactions, and required that
$B_2/B_{2hc}\ge-1$, then we would have obtained a distribution
of second-virial coefficients similar to that in Fig.~\ref{b2}.
In that sense a distribution like that in Fig.~\ref{b2} is generic
to any system where all model proteins except for those in
a large $B_2$ tail are cutoff.
However, within the earlier, simpler, model there is no clear way to look
at either mutations and hence evolution, or to compare with
sequence data for real proteins, as we did when we compared
Figs.~\ref{fhbac} and \ref{fh}.


Finally, many simplifying assumptions have been made in order to arrive at
our model system. It is therefore appropriate to comment on how
this work can be extended to include more of the features of proteins
inside cells. Both the model and our simple criterion for viability
can be
improved. The model is rather crude, and our sharp division between
proteins deemed soluble and those deemed insoluble, could be softened.
Then the fitness of a protein would decline over some range of
values of the second virial coefficient. Also, we did not impose
a maximum on the second virial coefficient. If it is important to limit
the osmotic pressure, values of the second virial coefficient which
are too positive may also be undesirable.
However, in terms of understanding
the behaviour of proteins in the complex crowded mixture of proteins
that is the {\em in vivo} environment, perhaps the most important
extensions of this work, is to multicomponent mixtures, and to include
proteins which bind to each other. Inside cells thousands of
different proteins
are mixed together at a total protein concentration of around 20\%,
and many proteins are not monomeric but are part of complexes.
The model studied here is flexible enough to both generate thousands
of different proteins and to permit selective binding between proteins.
Work on both is ongoing.

It is a pleasure to acknowledge that this work started
with inspiring discussions with D. Frenkel.
This work was supported by the Wellcome Trust (069242).

\section*{Appendix: Computations}

We are principally interested in the fraction
of proteins that are soluble according to our criterion,
and the distribution functions and means
of various properties of soluble proteins.
The fraction of proteins with
$B_2/B_{2hc}\ge -1$
is determined by simply generating a very large number of proteins at random
and finding the fraction that satisfy this requirement. The length
of all runs are determined either by the requirement to obtain at least
2 significant figures or until longer runs produce almost identical
plots. An exception is for $n_B=18$ and $\epsilon=2$ where
due to the smallness of $f_v$, it was only possible to obtain
1 significant figure of accuracy.
The distribution functions, means etc., are obtained by starting with
a soluble protein and generating a random walk in the space of soluble
proteins.
This is essentially no different from Metropolis Monte Carlo as applied to
a system with a hard potential, e.g., a fluid of hard spheres,
as our constraint $B_2/B_{2hc}\ge -1$, is a hard constraint.
The averages are then obtained over these random walks.

The algorithm samples
`protein space' \cite{smith70}, with each sequence a unique
point in this space and 2 sequences neighbours if 1 of them
can be transformed into the other by a single mutation.
This protein space is vast for real proteins and still very large
for our model; for our model it contains $2^{6n_B}$ points.
Note that {\em all} viable proteins
are connected to all other viable proteins by an unbroken path of
viable proteins and links between neighbouring viable proteins.
This is easy to see if we consider that $B_2$ always either increases
or stays the same if we flip a hydrophobic bit. Thus, starting
from any viable protein we can flip each of its hydrophobic bits
to hydrophilic bits, one at a time, until we reach the protein with
all $6n_B$ bits hydrophilic. Each intermediate in this path must
satisfy our solubility criterion as it is obtained from a protein
which satisfies this criterion by flipping 1 or more hydrophobic
bits. Thus we have proved that {\em all} viable proteins are connected
to the protein with all hydrophilic bits, and so trivially all
viable proteins are part of a connected network. This immediately
implies that we can go from any one
viable protein to any other via our Monte Carlo moves.


\begin{thebibliography}{99}
\bibitem{proteins} The two proteins have SWISSPROT \cite{swissprot,spref}
accession numbers Q9KTB7 and O66490. They are almost certainly descended
from a common ancestral protein and have, despite their very
different sequences, very similar 3-dimensional structures.
They are both part of the same family of proteins with accession
number PS00113. See the PROSITE \cite{prosite,prositeref}
database of protein families.
\end{thebibliography}
\end{document}